\documentclass{article}
\usepackage{spconf,amsmath,graphicx}
\usepackage{multirow,balance}
\usepackage{url}

\title{HLT-NUS SUBMISSION FOR 2020 NIST Conversational Telephone Speech SRE}
\name{Rohan Kumar Das, Ruijie Tao and Haizhou Li
}
\address{Department of Electrical and Computer Engineering,\\ National University of Singapore, Singapore}
\begin{document}
%
\maketitle
\begin{abstract}
This work provides a brief description of Human Language Technology (HLT) Laboratory, National University of Singapore (NUS) system submission for 2020 NIST conversational telephone speech (CTS) speaker recognition evaluation (SRE). The challenge focuses on evaluation under CTS data containing multilingual speech. The systems developed at HLT-NUS consider time-delay neural network (TDNN) x-vector and ECAPA-TDNN systems. We also perform domain adaption of probabilistic linear discriminant analysis (PLDA) model and adaptive s-norm on our systems. The score level fusion of TDNN x-vector and ECAPA-TDNN systems is carried out, which improves the final system performance of our submission to 2020 NIST CTS SRE. 
\end{abstract}
\section{Introduction}
\label{sec:intro}

The NIST speaker recognition evaluations\footnote{https://sre.nist.gov/} (SREs) are held annually/biannually that benchmark the systems developed at various sites with the latest technologies using a common test set. The current edition 2020 NIST SRE focuses on evaluating performance for conversational telephone speech (CTS) that contains multilingual speech unlike use of a particular language in the previous editions~\cite{SRE2020}. In addition, the challenge considers an open training condition and there are no separate development set released, which makes the challenge interesting among the participants as one have to create own development set for the studies. This work reports the systems developed HLT-NUS for challenge participation and the associated results.

\begin{table*}[t]
	\centering
	\caption{Description of various corpora considered for training TDNN x-vector and ECAPA-TDNN systems.}
	\vspace{1mm}
	{\begin{tabular}{|c|c||c|c|} \hline 
			{\bf Notation} &  {\bf Corpus}      & {\bf Notation} &{\bf Corpus} \\ \hline \hline
			SWB        & Switchboard			   & SRE          & SRE 2004-2016\\ \hline
			FSR       & Fisher                & Mx6   	  & Mixer 6 \\ \hline
			VoxC1      & VoxCeleb1		&   VoxC2       & VoxCeleb2\\ \hline
			SRE18      & SRE18 CTS EVAL		&   SRE19       & SRE19 CTS\\ \hline\hline
			{\bf TDNN x-vector System} & SWB+SRE+FSR+Mx6+VoxC1+VoxC2                 & {\bf ECAPA-TDNN System}       &  VoxC2+SRE18+SRE19     \\   \hline
		\end{tabular}}
		\label{data_part}
	\end{table*}

\section{TDNN x-vector System}

This section describes about the development of the time-delay neural network (TDNN) x-vector system developed for our studies. We develop our own system, using Kaldi\footnote{https://kaldi-asr.org} recipe \verb|egs/sre16/v2|. The details of our TDNN x-vector system are discussed in the following subsections.

\subsection{Data Resources}

As the challenge considers open training condition, it is important to have a strong training set for the studies covering wide range of conditions for CTS data. We consider databases from Linguistic Data Consortium (LDC) and a few open source databases like VoxCeleb to create the training set. The details of the databases chosen for training the TDNN x-vector system are shown in Table~\ref{data_part}.

\subsection{Features}

We consider mel frequency cepstral coefficient (MFCC) as feature for developing our system. For every utterance, 23-dimensional MFCC features are extracted considering short-term processing by using Hamming windowed frame of 25ms with a shift of 10ms. The feature extraction involves 23 mel filetbanks. In addition, the frames with sufficient voice activity are retained by performing an energy based VAD. 

\subsection{Front-end}

We perform data augmentation on the training dataset, which comprise of previous editions of SRE datasets 2004-2016, Mixer6, Switchboard, Fisher and VoxCeleb1-2 datasets. The number of training utterances are doubled by adding noisy and reverberated versions of the clean utterances. For this purpose, music, speech and babble noise segments extracted from the MUSAN database is used~\cite{MUSAN}. The MFCC features of the augmented training set are extracted for training a TDNN model based on the architecture described in~\cite{Snyder2017embedding,xvectors} to extract 512-dimensional x-vectors.

\subsection{Back-end}

The back-end of the system considers a 150-dimensional linear discriminant analysis (LDA) to reduce the dimension of the x-vectors followed by probabilistic linear discriminant analysis (PLDA) classifier to compute the likelihood scores. It is noted that the PLDA model is first adapted with SRE 2018 eval set before scoring, which we found to give a better result than that without domain adaptation. Additionally, score normalization is applied with adaptive s-norm technique considering the SRE 2018 eval set (top 30\% scores).

\section{ECAPA-TDNN System}

This section describes the details of the development of the emphasized channel attention, propagation and aggregation in time-delay neural network (ECAPA-TDNN) system~\cite{desplanques2020ecapa}. The training process contains the pretrain stage and the fine-tune stage. 

\subsection{Pretrain Stage}
The code of ECAPA-TDNN in this pretrain stage has been made available in GitHub.\footnote{https://github.com/TaoRuijie/ECAPATDNN}

\subsubsection{Data Resources and Augmentation}

We use the VoxCeleb2 dataset as the training data in the pretrain stage~\cite{Voxceleb2}. An online augmentation strategy is applied for data augmentation with RIR and MUSAN dataset~\cite{RIRS,MUSAN}. The augmentation methods include adding one kind of RIR, adding one music file, adding one noise file, adding three to eight speech files, adding one music file and one noise file. In addition, SpecAugment is applied with a frequency and temporal masking dimension of 8 and 10, respectively~\cite{SpecAug}. 

\subsubsection{Feature}

The duration of the input segment is 2 seconds. 80-dimensional log Mel-filterbank energies are extracted with a window length of 25ms and a frame-shift of 10ms as the input feature. After SpecAugment, the input features are mean normalized across time and fed into the speaker encoder.

\subsubsection{Front-end}

The speaker encoder is the standard ECAPA-TDNN structure~\cite{desplanques2020ecapa}, which contains 3 SE-Res2Block modules. The channel size and the dimension of the bottleneck in the SE-Block are set to 1024 and 256, respectively. Finally, the 192-dimensional speaker embedding can be extracted. We use AAM-softmax loss~\cite{deng2019arcface} and set margin and scale as 0.2 and 30, respectively.

\subsubsection{Back-end}

The entire utterance is fed into the speaker encoder in the evaluation process to obtain the speaker embedding. Then cosine similarity is used as the back-end to compute the score. The mean score between the utterances of the speaker is used as the final result.

\subsection{Implementation}

The initial learning rate is 0.001 that decreases 3\% in every one epoch. The mini-batch size is set at 400. The network parameters are optimized by Adam optimizer~\cite{2015Adam}.

\subsection{Fine-tune Stage}

Then we fine-tune the pretrain model on SRE18 CTS evaluation set and SRE19 CTS set. Here we up-sample all the utterances rate into 16 kHz. In this stage, we extend the duration of the input segments to 9 seconds. The margin of AAM-softmax loss is set as 0.5. The initial learning rate is 0.00005 that decrease 5\% in every epoch. The rest settings are all the same as those in the pretrain stage.

\section{Results and Analysis}

\subsection{System Performance}

Table~\ref{result} shows the performance of our systems and their score level fusion on SRE18 development set and SRE20 CTS progress set. It is observed that the ECAPA-TDNN system has a lower minDCF than the TDNN x-vector system. We performed score level fusion of both the systems to utilize the complementary information from them. The score fusion is carried out by BOSARIS toolkit~\cite{brummer2013bosaris} with the manually selected offset. However, we notice that the calibration in SRE20 CTS leads to a large difference between actDCF and minDCF. Hence, there is a large space to improve the actDCF.

\begin{table} [t!]
\caption{Performance on SRE18 Dev and SRE20 CTS progress set. }
\vspace{2mm}
\label{result}
\resizebox{8.7cm}{!}{
\begin{tabular}{|c|c|c|c|c|}
\hline
{\bf Dataset} & {\bf System} &  {\bf EER (\%)} & {\bf minDCF} & {\bf actDCF}\\
\hline\hline
\multirow{3}{*}{SRE18 Dev} & TDNN x-vector & 6.21 & 0.475 & - \\
                           & ECAPA-TDNN    & 6.76 & 0.389 & - \\\cline{2-5}
                           & Fusion        & 4.91 & 0.317 & 0.323 \\
\hline\hline
\multirow{3}{*}{SRE20 CTS}     & TDNN x-vector & 5.36 & 0.269 & -\\
                           & ECAPA-TDNN    & 5.35 & 0.239 & - \\\cline{2-5}
                           & Fusion        & 4.53 & 0.190 & 0.343\\
\hline
\end{tabular}
}
\vspace{-4mm}
\end{table}

\subsection{Hardware Description and Timing}

The experiments were performed on an Intel(R) Xeon(R) CPU E5-2630 v3 @ 2.40GHz CPU and an GeForce GTX TITAN X GPU. Table~\ref{tab:dev_datasets} shows the real time factors of all sub-systems for a single trial.

          \begin{table}[!th]
            \caption{\label{tab:dev_datasets} {Real time factor for processing a single trial.}}
            \vspace{2mm}            
            \centerline{
                          \begin{tabular}{|c|c|}
                \hline
                 {\bf System} & {\bf Real Time Factor}  \\ \hline\hline
                 TDNN x-vector & 0.02  \\ \hline
                 ECAPA-TDNN & 0.05  \\ \hline
              \end{tabular}}
             \end{table}  

\section{Conclusion}

In this work, we reported the systems developed by HLT-NUS for 2020 NIST CTS SRE Challenge. Our systems comprise of two different frameworks, which are TDNN x-vector and ECAPA-TDNN system. Both the systems use different set of training data as the challenge focuses on open training conditions. We use domain adaption of the PLDA model and perform adaptive s-norm on our systems. A score level fusion of the two systems improves the system performance due to complementary nature of information from both systems.

\balance
\bibliographystyle{IEEEbib}
\bibliography{MyReferences_new}

\end{document}